\documentclass[preprint,aps,nofootinbib]{revtex4-1}
\pdfoutput=1

\usepackage{amsmath}  
\usepackage{amsbsy,amssymb,amsfonts}
\usepackage{graphicx}
\usepackage{color}
\usepackage{epsf}
\usepackage[utf8x]{inputenc}
\usepackage{hyperref}
\usepackage{multirow}
\def\bravert{\egroup\,\vrule\,\bgroup}

{\catcode`\|=\active
  \gdef\Twoint#1{\left(\mathcode`\|"8000\let|\bravert {#1}\right)}}
{\catcode`\|=\active
  \gdef\Braket#1{\left<\mathcode`\|"8000\let|\bravert {#1}\right>}}

\newcommand{\beq}{\begin{equation}}
\newcommand{\eeq}{\end{equation}}
\newcommand{\beqa}{\begin{eqnarray}}
\newcommand{\eeqa}{\end{eqnarray}}
\newcommand{\bea}{\begin{array}}
\newcommand{\eea}{\end{array}}

\newcommand{\bef}{\begin{figure}}
\newcommand{\ef}{\end{figure}}
\newcommand{\bc}{\begin{center}}
\newcommand{\ec}{\end{center}}
\newcommand{\bt}{\begin{table}}
\newcommand{\et}{\end{table}}
\newcommand{\btb}{\begin{tabular}}
\newcommand{\etb}{\end{tabular}}

\newcommand{\au}{{\em a.u.}}

\def\rvac{\left| \rule{0.3cm}{.0cm} \right>}

\def\etal{{\it et al.\ }}
\def\au{{\it a.u.\ }}
\begin{document}

\title { ${\cal{P,T}}$-Odd Weak Neutral Current Interactions in the TlF Molecule }

\vspace*{1cm}

\author{Timo Fleig}
\email{timo.fleig@irsamc.ups-tlse.fr}
\affiliation{Laboratoire de Chimie et Physique Quantiques,
             FeRMI, Universit{\'e} Paul Sabatier Toulouse III,
             118 Route de Narbonne, 
             F-31062 Toulouse, France }
\vspace*{1cm}
\date{\today}

\vspace*{1cm}
\begin{abstract}
Leptoquark models may explain deviations from the Standard Model observed in decay processes involving heavy quarks at 
high-energy colliders. Such models give rise to low-energy 
parity- and time-reversal-violating phenomena in atoms and molecules. One of the leading effects among these phenomena is the 
nucleon-electron tensor-pseudotensor interaction when the low-energy experimental probe uses a quantum state of an atom or 
molecule predominantly characterized by closed electron shells.

In the present paper the molecular interaction constant for the nucleon-electron tensor-pseudotensor interaction in the 
thallium-fluoride molecule -- used as such a sensitive probe by the CeNTREX collaboration [Quantum Sci. Technol., 6:044007, 2021]
-- is calculated employing highly-correlated relativistic many-body theory. Accounting for up to quintuple excitations in the
wavefunction expansion the final result is $W_T({\text{Tl)}} = -6.25 \pm 0.31\, $[$10^{-13} {\langle\Sigma\rangle}_A$ \au]
Interelectron correlation effects on the tensor-pseudotensor interaction are studied for the first time in a molecule, and
a common framework for the calculation of such effects in atoms and molecules is presented.

\end{abstract}

\maketitle
\section{Introduction}
\label{SEC:INTRO}
In the last decade tests of observables in decays involving $b$ ({\it{beauty, bottom}}) quarks have shown deviations from
the predictions offered by the Standard Model (SM) of elementary particles 
(\cite{LeptonUniversality_2023,lepton_univ_viol2022,AngularAnalysis_BK_decay_2021} and references therein).
These observed deviations could be explained through the existence of New Physics (NP) particles, the leptoquarks 
\cite{Becirevic:2018afm,Barbieri_compLeptoquark_2017,Wise_leptoquarks2013,Davies_He_scalarfermions1990,Buchmueller_leptoquarks1987},
among other possible models.

Electric dipole moments (EDMs) of atoms and molecules are highly sensitive low-energy probes 
\cite{https://doi.org/10.48550/arxiv.2203.08103} of charge-parity (${\cal{CP}}$)-violating physics beyond that already 
implemented in the Standard Model (SM) elementary particles.
Under the assumption that the considered atomic or molecular quantum state has predominantly closed electron shells,
scalar leptoquark models with a dominant contribution through light quarks amplify 
\cite{Barr_eN-EDM_Atoms_1992,He_McKellar_eNinteractions_1992} a 
${\cal{CP}}$-violating (CPV) tensor-pseudotensor nucleon-electron (T-PT-ne) 
interaction over other possible sources of an atomic-scale EDM, with the exception \cite{Cairncross_Ye_NatPhys2019}
of a collective nuclear effect, the nuclear Schiff moment \cite{ginges_flambaum2004}.
By contrast, in open-shell atomic or molecular states the contribution due to
the CPV scalar-pseudoscalar nucleon-electron (S-PS-ne) interaction is the leading one among the
possible nucleon-electron four-fermion interactions. 
In other words, EDM measurements on closed-shell atomic or molecular systems
with a very high sensitivity to underlying ${\cal{CP}}$-violating NP probe such scalar leptoquark models. This is achieved
by interpreting the measured EDM -- or measured upper bound on an EDM -- in terms of the ${\cal{CP}}$-odd T-PT-ne 
parameter $C_T$ and an atomic or molecular interaction constant that is calculated by relativistic many-body theory.

The complex system delivering the most sensitive low-energy probe of light-quark scalar leptoquark models at the moment
is the mercury (Hg) atom \cite{Heckel_Hg_PRL2016}. Efforts are currently undertaken to significantly improve upon the 
sensitivity of the Hg EDM measurement by using the thallium-fluoride (TlF) molecule in a cold molecular-beam experiment 
(CeNTREX collaboration, \cite{CENTREX_2021}). This molecular EDM measurement on the electronically closed-shell TlF
ground state can be interpreted \cite{Hubert_Fleig_2022} in terms of the nuclear Schiff moment of the {$^{205}$Tl} nucleus
and in terms of more fundamental CPV parameters, the strong $\pi$-meson exchange constants, the QCD
$\Theta$ parameter, chromo-EDMs and neutron- and proton EDMs \cite{Flambaum-Dzuba_TranPRA2020}. In addition, the stable
isotope {$^{205}$Tl} has nuclear spin quantum number $I = \frac{1}{2}$ \cite{Proctor_nuclear1950,Bounds_Tl_nuclear1987} 
which makes it sensitive to the nuclear-spin-dependent T-PT-ne interaction. 

However, {$^{205}$Tl} is not located in the region of strongly octupole-deformed nuclei \cite{Budincevic_diss}. Indeed, it has
been shown \cite{Flambaum-Dzuba_PRA2020,Flambaum-Dzuba_TranPRA2020} that the dependency of the Schiff moment of 
{$^{205}$Tl} on underlying CPV parameters such as QCD $\Theta$ and the $\pi$-meson exchange constants 
is some orders of magnitude smaller than same dependency of the Schiff moments of {$^{225}$Ra} or {$^{223}$Fr$^+$}. This 
relatively feeble dependency is also established for {$^{199}$Hg}. Thus, the interpretation of the TlF EDM in terms of the molecular 
T-PT-ne interaction in that system may even be more important than in terms of the Schiff moment of its {$^{205}$Tl} nucleus. 

The earliest calculation of the T-PT-ne interaction constant in a molecule has been reported by Coveney and Sandars in 1983
as it happens for the TlF molecule \cite{Coveney_JPB1983}. Dirac-Hartree-Fock calculations by Quiney 
{\etal} in 1998 \cite{Quiney_PTodd_PRA1997} gave a result more than $500$\% greater than the earlier calculation.
Since then, no calculations of T-PT-ne interaction constants have been reported in any molecule.

The purpose of the present paper is twofold: First, the theory of atomic and molecular T-PT-ne interactions is reviewed and
corresponding interaction constants are defined in a common framework for atoms and molecules (section \ref{SEC:THEORY}). 
Such a common framework is a distinct advantage since it eliminates problems arising from differing
conventions in global analyses of ${\cal{CP}}$-violation effects \cite{Chupp_Ramsey_Global2015,FleigJung_JHEP2018}.
Second, the T-PT-ne interaction constant of ground-state TlF is calculated by relativistic many-body
theory of general excitation rank in section \ref{SEC:RESULTS}. The present results both resolve the massive discrepancy between
earlier calculations and also provide a value that includes interelectron correlation effects to high accuracy. Correlation
effects on the T-PT-ne interaction are presented for the first time in a molecule. In the final section (\ref{SEC:CONCL}) 
the expected impact of the present findings is discussed in particular considering future measurements of the TlF EDM.

\section{Theory}
\label{SEC:THEORY}
\subsection{Particle physics}
\begin{figure}[h]
  \includegraphics[angle=0,width=9.0cm]{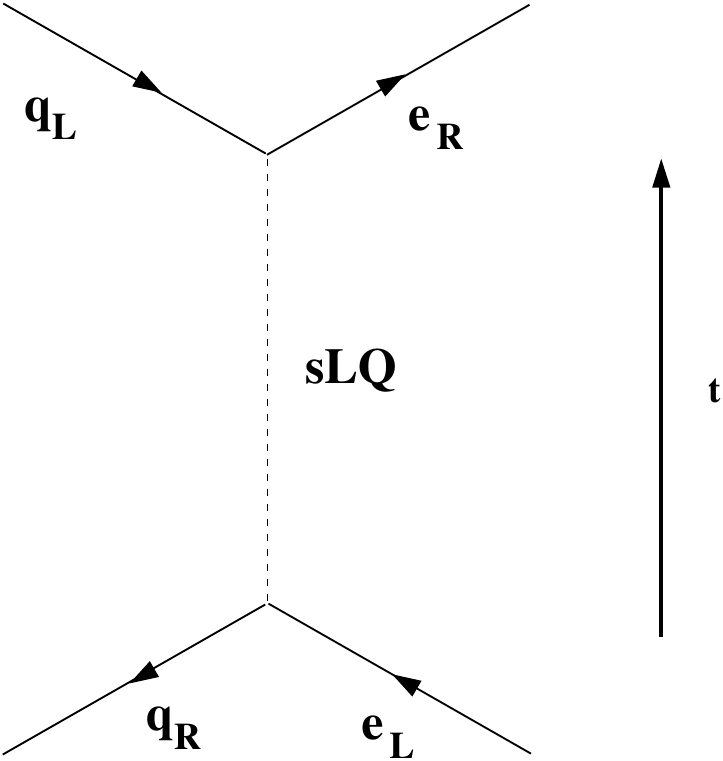}

  \medskip
  \caption{Tree-level diagram for the exchange of a scalar (spin-0) leptoquark (sLQ) between an anti-quark and an electron; 
           this process resembles Bhabha scattering in quantum electrodynamics when sLQ $\longrightarrow \gamma$ and
           $\overline{q} \longrightarrow \overline{e}$; R/L stand for fermion chirality.}
  \label{FIG:SLQEX}
\end{figure}

As Barr expounds in Ref. \cite{Barr_eN-EDM_Atoms_1992} the LQ exchange displayed in Fig. (\ref{FIG:SLQEX}) leads to
${\cal{P}}$- and ${\cal{T}}$-odd nucleon-electron four-fermion coupling coefficients $C_{S(q)}, C_{P(q)}$ and $C_{T(q)}$ at the
quark $(q)$ level which are all non-vanishing. $S$ here denotes nucleon-electron scalar-pseudoscalar (S-PS), $P$ nucleon-electron 
pseudoscalar-scalar (PS-S) and $T$ nucleon-electron tensor-pseudotensor (T-PT) coefficients, respectively. 
 
If light quarks ($u$, $d$) make the dominant contribution to the ${\cal{P}}$- and ${\cal{T}}$-odd couplings in such models
then the ratio of the T-PT and S-PS coefficients is found \cite{Barr_eN-EDM_Atoms_1992} to be $\frac{C_T}{C_S} \approx
\frac{1}{10}$. This suppression of the T-PT contribution by about one order of magnitude relative to the S-PS contribution
to the ${\cal{P}}$- and ${\cal{T}}$-odd effects, however, only holds in a complex system if the corresponding interaction
constants are of the same order of magnitude. Conversely, a suppression of the S-PS contribution to a complex-system EDM
can be achieved by using a quantum state with predominantly paired (closed) electron shells where S-PS contributions
cancel pairwise in lowest order of perturbation theory.

\subsection{Isotope-specific effective many-electron T-PT-ne interaction Hamiltonian}

The T-PT-ne atomic interaction constant can be defined starting from an effective theory at the level of nucleons \cite{Barr_eN-EDM_Atoms_1992}.
%
An effective first-quantized Hamiltonian for a single electron can be written as \cite{Fleig_Jung_Xe_2021}
\begin{equation}
 \label{EQ:T-PT_HAM1}
 \hat{H}^{\text{eff}}_{\text{T-PT-ne}} = \frac{\imath G_F}{\sqrt{2}}\,\sum_N C_T^N\, \rho_N({\bf{r}})
                \gamma^0 {\sigma_N}_{\mu\nu} \gamma^5 {\sigma}^{\mu\nu}\,,
\end{equation}
where $\rho_N({\bf{r}})$ denotes the probability density of the corresponding nucleon at the position of the electron
and the electronic Dirac matrix $\gamma^0$ originates in the field creator ${\overline{\psi}}$.
$C_T^N$ are Wilson coefficients determined by ${\cal{CP}}$-odd interactions at higher energies.
The term ${\sigma_N}_{\mu\nu} \gamma^5 {\sigma}^{\mu\nu}$ satisfies the following identity:
\begin{equation}
 \label{EQ:T-PT_EXPRESSION}
 {\sigma_N}_{\mu\nu} \gamma^5 {\sigma}^{\mu\nu} = 2 \gamma_N^0 {\boldsymbol{\gamma}}_N \cdot {\boldsymbol{\Sigma}}
                                                + 2 \gamma^0 {\boldsymbol{\Sigma}}_N \cdot {\boldsymbol{\gamma}}\,,
\end{equation}
where ${\boldsymbol{\Sigma}}_N$ denotes the nuclear and ${\boldsymbol{\Sigma}}$ the electronic spin matrix. 
In Ref. \cite{Fleig_Jung_Xe_2021} it was shown that the electronic expectation value 
$\left<\psi | {\boldsymbol{\Sigma}} | \psi\right>$ of the first term on the right-hand side of Eq. (\ref{EQ:T-PT_EXPRESSION}) 
is strictly zero if $\psi$ is a closed-shell wavefunction for a many-electron state with valence configuration $ns_{1/2}^2$.
The T-PT-ne Hamiltonian used in the present is then written as
\begin{equation}
 \label{EQ:T-PT_HAM_FIN}
 \hat{H}^{\text{eff}}_{\text{T-PT-ne}} = \frac{\imath G_F}{\sqrt{2}}\, \sum_N 2 C^T_N\, {\boldsymbol{\Sigma}}_N \cdot 
  {\boldsymbol{\gamma}} \rho_N({\bf{r}})\,.
\end{equation}
In terms of nuclear spin $I$, $\langle\mathbf \Sigma\rangle = \langle\Sigma\rangle\, \mathbf I/I$. Introducing a nuclear
isotope-specific ${\cal{CP}}$-odd parameter $C_T^A$ Eq. (\ref{EQ:T-PT_HAM_FIN}) becomes
\begin{equation}
 \hat{H}^{\text{eff}}_{\text{T-PT-ne}} = \frac{2\imath G_F}{\sqrt{2}}\,  \rho(\mathbf r)\,C_T^A\,{\langle\Sigma\rangle}_A \frac{\mathbf I\cdot \boldsymbol{\gamma}}{I}\,.
\end{equation}
where $\rho(\bf r)$ now refers to a nuclear density assumed to be equal for protons and neutrons.

In a setup with rotational symmetry around the $z$ axis, the $1,2$ components of $\langle\mathbf \Sigma\rangle$ vanish. 
Considering furthermore a nuclear state $|I,M_I=I\rangle$ and integrating over the nuclear coordinates, the Hamiltonian can 
be written as
\begin{equation}
 \hat{H}^{\text{eff}}_{\text{T-PT-ne}} = \imath\sqrt{2} G_F\, C_T^A\,{\langle\Sigma\rangle}_A \gamma_3 \,  \rho(\mathbf r)\,.
 \label{EQ:HEFF_SINGLE}
\end{equation}
On the basis of this effective single-electron Hamiltonian the molecular T-PT-ne interaction can be defined. In an $n$-electron
framework, Eq. (\ref{EQ:HEFF_SINGLE}) becomes
\begin{equation}
 \hat{H}^{\text{eff}}_{\text{T-PT-ne}} = \imath\sqrt{2} G_F\, C_T^A\,{\langle\Sigma\rangle}_A \sum\limits_{j=1}^n\, (\gamma_3)_j \,  \rho(\mathbf r_j)\, .
 \label{EQ:HEFF_MANY}
\end{equation}

\subsection{Molecular T-PT-ne Interaction Constant}

The T-PT-ne interaction is evaluated through lowest order in perturbation theory. The wavefunction of an electronic state $E$
is the approximate solution of an eigenvalue problem
\begin{equation}
 \hat{H}^{DC} \psi^{(0)}_E = \varepsilon_E\, \psi^{(0)}_E
\end{equation}
where $\hat{H}^{DC}$ is the Dirac-Coulomb Hamiltonian
\begin{equation}
 \hat{H}^{DC} = \sum\limits^n_j\, \left[ c\, \boldsymbol{\alpha}_j \cdot {\bf{p}}_j + \beta_j c^2
                - \sum\limits^2_K\, \frac{Z_K}{r_{jK}}{1\!\!1}_4 \right]
                + \sum\limits^n_{k>j}\, \frac{1}{r_{jk}}{1\!\!1}_4 + V_{KL}
 \label{EQ:HAMILTONIAN_MOL}
\end{equation}
for a diatomic molecule with $n$ electrons,
where $\boldsymbol{\alpha}, \beta$ are electronic Dirac matrices, $K$ runs over nuclei and $V_{KL}$ is the classical 
electrostatic potential energy for the two Born-Oppenheimer-fixed nuclei. 

$\psi^{(0)}_E$ is linearly expanded
\begin{equation}
        \left| \psi^{(0)}_E \right> = \sum\limits_{L=1}^{{\rm{dim}}{\cal{F}}^t(M,n)}\,
                                       c^E_{(M_J),L}\, ({\cal{S}}{\overline{\cal{T}}})^E_{(M_J),L} \rvac
        \label{EQ:MOL_WF}
\end{equation}
where $\rvac$ is the true vacuum state,
${\cal{F}}^t(M,n)$ is the symmetry-restricted sector of Fock space ($M_J$ subspace with $J$ total electronic angular momentum) 
with $n$ electrons in $M$ four-spinors,
${\cal{S}} = a^{\dagger}_i a^{\dagger}_j a^{\dagger}_k \ldots$ is a string of spinor creation operators,
${\overline{\cal{T}}} = a^{\dagger}_{\overline l} a^{\dagger}_{\overline m} a^{\dagger}_{\overline n} \ldots$
is a string of creation operators of time-reversal transformed spinors. The determinant expansion coefficients
for state $E$,
$c^E_{(M_J),L}$, are obtained through relativistic general-excitation-rank Configuration Interaction theory as detailed
in Refs. \cite{fleig_gasci,fleig_gasci2}.

The energy shift of state $E$ in a molecule due to the T-PT-ne interaction is now given as an expectation value 
over the effective Hamiltonian in Eq. (\ref{EQ:HEFF_MANY})
\begin{equation}
 \Delta\varepsilon_E = \left< \psi^{(0)}_E \right| \hat{H}^{\text{eff}}_{\text{T-PT-ne}} \left| \psi^{(0)}_E \right>
                     = W_T\, C_T^A
 \label{EQ:EXPVAL}
\end{equation}
and expressed in terms of a molecular interaction constant $W_T$ and the effective ${\cal{CP}}$-odd parameter $C_T^A$. 
It then follows from Eqs. (\ref{EQ:HEFF_MANY}) and (\ref{EQ:EXPVAL}) that the molecular T-PT-ne interaction constant is
\begin{equation}
  W_T(X) = \sqrt{2} G_F {\langle\Sigma\rangle}_A \, 
           \left< \psi^{(0)}_E \left| \imath \sum\limits_{j=1}^n\, \left(\gamma_3\right)_j\, \rho_X({\bf{r}}_j) \right| \psi^{(0)}_E \right>
  \label{EQ:WTDEF}
\end{equation}
where $X$ denotes the individual atomic nucleus in the molecule.
The expectation value $\left<\psi | \gamma_3 | \psi\right>$ for a closed-shell electronic state is shown to be non-zero in Ref.
\cite{Fleig_Jung_Xe_2021}. $W_T$ has dimension of energy.

For purposes of comparison the atomic case \cite{Fleig_PRA2019} is briefly reviewed.
The atomic electric dipole moment due to a tensor-pseudotensor interaction can be written as
\begin{equation}
 d_a = C_T^A\, \alpha_{C_T}
\end{equation}
where 
\begin{equation}
  \alpha_{C_T} := \frac{\left< \hat{H}^{\text{eff}}_{\text{T-PT-ne}} \right>_{\psi^{(0)}(E_{\text{ext}})}}{E_{\text{ext}}}
  \label{EQ:ALPHA_CT}
\end{equation}
and ${\psi^{(0)}}(E_{\text{ext}})$ is the field-dependent atomic wavefunction of the state in question.
$\alpha_{C_T}$ has physical dimensions of an electric dipole moment and is evaluated in the quasi-linear regime 
(see also Ref. \cite{Fleig_Skripnikov2020}) with very small external electric fields $E_{\text{ext}}$.

\section{Tensor-pseudotensor interaction in molecules}
\label{SEC:RESULTS}
\subsection{Consistency test}

A rigorous test of the molecular interaction constant is here carried out by calculating the atomic T-PT-ne interaction in
the Xe atom in two different ways:
\begin{enumerate}
  \item A Xe atom is subjected to a finite external electric field along the $z$ axis and $\alpha_{C_T}$ is calculated using
        Eq. (\ref{EQ:ALPHA_CT}). In a cvTZ basis set this yields a DCHF value of
        $\alpha_{C_T} = 1.0422 \times 10^{-12} {\langle\Sigma\rangle}_A$ \au
  \item A Xe atom is placed in the electric field of an F$^-$ ion at a distance of $40$ \au and $W_T$ is calculated through
        Eq. (\ref{EQ:WTDEF}) in a molecular calculation. De facto, this corresponds to the calculation of the interaction
        constant in a stretched XeF$^-$ ion. The electric field of the F$^-$ ion at the position of the Xe atom is then
        $E({\text{F}}^-) = 0.000625$ \au  Through calculation,
        $\frac{W_T({\text{Xe}})}{E({\text{F}}^-)} = 1.0417 \times 10^{-12} {\langle\Sigma\rangle}_A$ \au,
        confirming that the molecular evaluation is consistent with an atomic calculation. The slight difference
        between the results of the atomic and the molecular evaluation can be explained by the fact that in the atomic
        case the external field is homogeneous across the Xe atom whereas it is not perfectly homogeneous when created
        by an F$^-$ ion at a distance.
\end{enumerate}

\subsection{Tensor-pseudotensor interaction in TlF}

\subsubsection{Technical details}
Atomic Gaussian basis sets of Double-Zeta (DZ), Triple-Zeta (TZ) and Quadruple-Zeta (QZ) quality are used for both 
atoms, in the case of
Tl the Dyall basis sets \cite{dyall_p,dyallBi2,4p-basis-dyall-2} including valence-correlating, 5d-correlating, and 
4f-correlating exponents and in the case of F the cc-pVNZ sets from the EMSL library \cite{EMSL-basis2019}.
Electronic-structure calculations are carried out using a locally-modified version of the DIRAC program package
\cite{DIRAC_JCP}. For correlated calculations beyond DC Hartree-Fock (HF) the Kramers-Restricted Configuration
Interaction (KRCI) module \cite{knecht_luciparII,fleig_gasci2} is employed for obtaining relativistic molecular
wavefunctions and for the evaluation of the T-PT-ne interaction constants. For the case of Hartree-Fock theory
the ground-state Slater determinant comprises the wavefunction $\psi^{(0)}_E$ in Eq. (\ref{EQ:WTDEF}).
All single-point Born-Oppenheimer calculations are carried out at the experimental equilibrium interuclear distance 
of $R_e = 3.94$ \au\ \cite{Barrett_Mandel_1957}.

\subsubsection{Hartree-Fock theory}
In Table \ref{TAB:TLF_DCHF} I compare various results calculated through Hartree-Fock theory. Basis-set effects are
seen to be very small, and the largest basis used in the present work -- augmented for the calculation of Schiff-moment
interactions \cite{Hubert_Fleig_2022} -- does not improve upon the standard QZ basis set. For this reason the correlated
calculations in the following section are carried out with the QZ basis for both atoms. 

\begin{table}[h]
        \caption{\label{TAB:TLF_DCHF}
                 T-PT-ne interaction constant in TlF($^1\Sigma_0$) at $R = 3.94$ \au\ from Hartree-Fock theory
                }

  \vspace*{0.3cm}
  \begin{tabular}{l|cl}
          model       & $W_T({\text{Tl}})\,$ [kHz ${\langle\Sigma\rangle}_A$] & total energy       \\ \hline
          Hartree-Fock\footnotemark[1]       &        $-0.851$  &                                  \\
          Dirac-Coulomb HF\footnotemark[2]   &        $-4.330$  &   $-20374.4108$                  \\ \hline
          DZ/DCHF                            &        $-4.601$  &   $-20374.41122770$              \\
          TZ/DCHF                            &        $-4.673$  &   $-20374.46576781$              \\
          QZ/DCHF                            &        $-4.684$  &   $-20374.47704191$              \\
          QZ+dens+sp/DCHF                    &        $-4.684$  &   $-20374.47660904$  
  \end{tabular}
  \footnotetext[1]{Value reported by Cho {\etal} \cite{ChoSangsterHinds_TlF_PRA1991} which is the corrected result from Ref. 
                   \cite{Coveney_JPB1983}}
  \footnotetext[2]{Value from Ref. \cite{Quiney_PTodd_PRA1997} in converted units}
\end{table}

The present results agree with the 1991 results by Quiney \etal\ which have also been obtained in the framework
of Dirac theory. The most accurate present basis set increases the absolute result by Quiney \etal\ by about $7$\%. The latter
was obtained at an internuclear distance $R$ of $3.953$ \au, but correcting for the difference in $R$ merely changes
it to roughly $-4.35$ kHz. The total energies given in Table \ref{TAB:TLF_DCHF} and the trend for $W_T$ with basis
sets of increasing extent strongly suggest that the difference between the present results and that from Quiney \etal\
is due to the higher quality of Tl basis sets used in the present work. From the consistency of these results it is also 
evident that the early method employed by Coveney and Sandars in a framework that does not account for the
relativistic character of the $s-$ and $p-$type TlF spinors in a rigorous fashion produces quite large errors. Not
surprisingly, T-PT-ne interactions should be calculated in a fully relativistic atomic or molecular framework.

\subsubsection{Correlated theory}
Results for the T-PT-ne interaction constant from a systematic series of calculations including interelectron correlation
effects are compiled in Table \ref{TAB:TLF_DCCI}. Not surprisingly, the main correlation effect comes from the valence 
electrons in the Tl($6s$) and F($2p_z$) shells and amounts to about $-11$\% in magnitude (from model QZ/SDTQ4). 
Like other ${\cal{P,T}}$-odd effects in molecules \cite{Hubert_Fleig_2022,PhysRevA.95.022504} also the T-PT-ne interaction
is dominated by $s-p$ mixing in the valence shells of the heavy target atom.
Interestingly, correlations with and among the remaining valence electrons in the F($2p_{x,y}$) shells are not
negligible and result in another shift of $W_T$ of about $-3.3$\% in magnitude (model QZ/SDTQ8). Since full triple and 
full quadruple excitations among the $8$ valence electrons change $W_T$ by more than $5$\% (relative to QZ/SD8) it is not
unreasonable to assume that even higher excitations could result in further non-negligible corrections. This is,
however, not the case, as to be seen from the comparison of models QZ/S4\_SDTQ8 and QZ/S4\_SDTQQ8 which yield
almost identical results. In addition, the chosen virtual cutoff of $6.5$ \au\ for valence correlations is seen to be 
sufficiently accurate through comparison with results at $11$ \au 
The value of $-6.12$ [$10^{-13} {\langle\Sigma\rangle}_A$ \au] for $W_T$ can, therefore,
be considered as sufficienly converged as concerns valence-electron correlations.

\begin{table}[h]
        \caption{\label{TAB:TLF_DCCI}
         Interelectron correlation effects on the T-PT-ne interaction constant in TlF($^1\Sigma_0$) at $R = 3.94$ \au\
         and comparison with the literature; present models are denoted as ``Basis set/CI model/cutoff in virtual space'';
         A CI model denoted ``SD$m$\_SDTQ$n$'' means that single, double, triple and quadruple replacements out of shells with 
         $n$ electrons are performed where subshells with $m$ electrons are restricted to single and double replacements;
         $n=4$ corresponds to the Tl($6s$) and F($2p_z$) shells, $n=8$ corresponds to the Tl($6s$) and all F($2p$) shells, 
         $n=18$ adds Tl($5d$) to $n=8$, $n=20$ adds F($2s$) to $n=18$, 
         $n=26$ adds Tl($5p$) to $n=20$, $n=28$ adds Tl($5s$) to $n=26$; 
         $n=36$ adds Tl($4s,4p$) to $n=28$; 
         $m=4$ denotes the F($2p_{x,y}$) subshells. 
        }

 \vspace*{0.3cm}
 \hspace*{-1.0cm}
 \begin{tabular}{l|ccc}
                         & \multicolumn{2}{c}{$W_T({\text{Tl}})$} & \\
         model           & [$10^{-13} {\langle\Sigma\rangle}_A$ \au] & [kHz ${\langle\Sigma\rangle}_A$]  & total energy  \\ \hline
         QZ/DCHF                      & $-7.12$         &    $-4.68$                  & $-20374.47704191$  \\
         QZ/SD4\_6.5au                & $-6.40$         &    $-4.21$                  & $-20374.53183162$  \\
         QZ/SDTQ4\_6.5au              & $-6.33$         &    $-4.17$                  & $-20374.53321310$  \\ \hline
         QZ/S4\_SDTQ8\_6.5au          & $-6.08$         &    $-4.00$                  & $-20374.61641168$  \\
         QZ/SDTQ4\_SDTQ8\_6.5au       & $-6.12$         &    $-4.03$                  & $-20374.69523546$  \\
         QZ/S4\_SDTQQ8\_6.5au         & $-6.07$         &    $-4.00$                  & $-20374.61647310 $  \\ \hline
         QZ/SD8\_2au                  & $-6.27$         &    $-4.12$                  & $-20374.58577569$  \\
         QZ/SD8\_6.5au                & $-6.47$         &    $-4.26$                  & $-20374.67877868$  \\
         QZ/SD8\_11au                 & $-6.46$         &    $-4.25$                  & $-20374.67958679$  \\
         QZ/SDT8\_6.5au               & $-6.47$         &    $-4.26$                  & $-20374.68490950$  \\
         QZ/SDTQ8\_6.5au              & $-6.12$         &    $-4.03$                  & $-20374.69523546$  \\ \hline
         QZ/SD18\_6.5au               & $-6.58$         &    $-4.33$                  & $-20374.98617776$  \\ \hline
         QZ/SD20\_6.5au               & $-6.61$         &    $-4.35$                  & $-20375.05000858$  \\ \hline
         QZ/SD26\_6.5au               & $-6.55$         &    $-4.31$                  & $-20375.15736493$  \\ \hline
         QZ/SD28\_6.5au               & $-6.49$         &    $-4.27$                  & $-20375.17819744$  \\ 
         QZ/SD28\_11au                & $-6.57$         &    $-4.32$                  & $-20375.32222400$  \\
         QZ/SD28\_18au                & $-6.59$         &    $-4.33$                  & $-20375.37322100$  \\   
         QZ/SD28\_30au                & $-6.60$         &    $-4.34$                  & $-20375.39255661$  \\
         QZ/SD28\_50au                & $-6.60$         &    $-4.34$                  & $-20375.40305619$  \\ \hline
         QZ/SD36\_18au                & $-6.59$         &    $-4.34$                  & $-20375.38490094$  \\ \hline   
         {\bf{Final}}                 & {\bf{-6.25}}    &    ${\bf{-4.11}}$           &                    \\
  \end{tabular}
\end{table}

Effects due to core-valence and core-core interelectron correlations follow a familiar pattern \cite{Hubert_Fleig_2022} for
${\cal{P,T}}$-odd effects in closed-shell systems. When electrons from heavy-atom shells of $s-$ and $p-$type are involved in
excitations $W_T$ drops because the mean density in shells contributing directly to the ${\cal{P,T}}$-odd effect is reduced
(model SD26 relative to SD20 and SD28 relative to SD26). By contrast, when electrons are removed from occupied shells with
higher angular momentum $W_T$ increases (model SD18 relative to SD8) since in that case the mean density in shells 
contributing directly to the ${\cal{P,T}}$-odd effect is increased. These cancellations make the total correlation effect
from core shells down to Tl($5s$) quite small in total at the SD level, less than $20$\% of valence correlation effects.
This and the rising energy denominators for contributions from core spinors strongly suggests that correlation effects 
from deeper core shells will make significantly smaller contributions. Indeed, the model SD36 which includes correlations
from the Tl($4s,4p$) electrons changes $W_T$ by less than $0.1$\%. Similar insignificance can thus also be assumed for 
higher excitation ranks up to full quadruple excitations involving the core electrons.

The final value for $W_T$ is calculated as follows. The base value is comprised by the highest-level correlated calculation
with the greatest number of active electrons, model QZ/SD28\_50au. This value is corrected by adding correlation effects
from excitation ranks up to full quadruple excitations among the valence electrons. Formally, this means
\begin{equation}
  W_T({\text{\bf{Final}}}) = W_T({\text{QZ/SD28\_30au}}) + W_T({\text{QZ/SDTQ8\_6.5au}}) - W_T({\text{QZ/SD8\_6.5au}})
\end{equation}
The final value, listed in Table \ref{TAB:TLF_DCCI}, is thus $W_T = -6.25$ [$10^{-13} {\langle\Sigma\rangle}_A$ \au], 
showing that the total modification due to interelectron correlations is $-12.2$\% in magnitude. 
The statement by Quiney \etal\ (\cite{Quiney_PTodd_PRA1997}, p.939) that it were ``unlikely that electron correlation will 
have a significant effect on $\ldots T$ [the T-PT-ne interaction constant]'' was evidently premature.

The uncertainty in the present final result is estimated from an unweighted sum over individual uncertainties for effects
contributing to the T-PT interaction. An individual uncertainty is defined as the difference between the most accurate and
the second most accurate model used in the present work for a given effect. This results in individual uncertainties of
$1$\% for atomic basis sets, $1$\% for valence-electron correlations, $1$\% for core-correlation effects, and $1$\% for the
cutoff in the virtual spinor space. To this are added $1$\% for neglecting the Breit and radiative corrections 
\cite{singh_CPodd-Ra_PRA2015}. The final $5$\% is a conservative uncertainty estimate since partial cancellations among the 
different neglected contributions are very likely.

\subsubsection{$W_T$ as a function of internuclear distance}

\begin{figure}[h]

  \hspace*{-2.5cm}
  \includegraphics[angle=0,width=22.0cm]{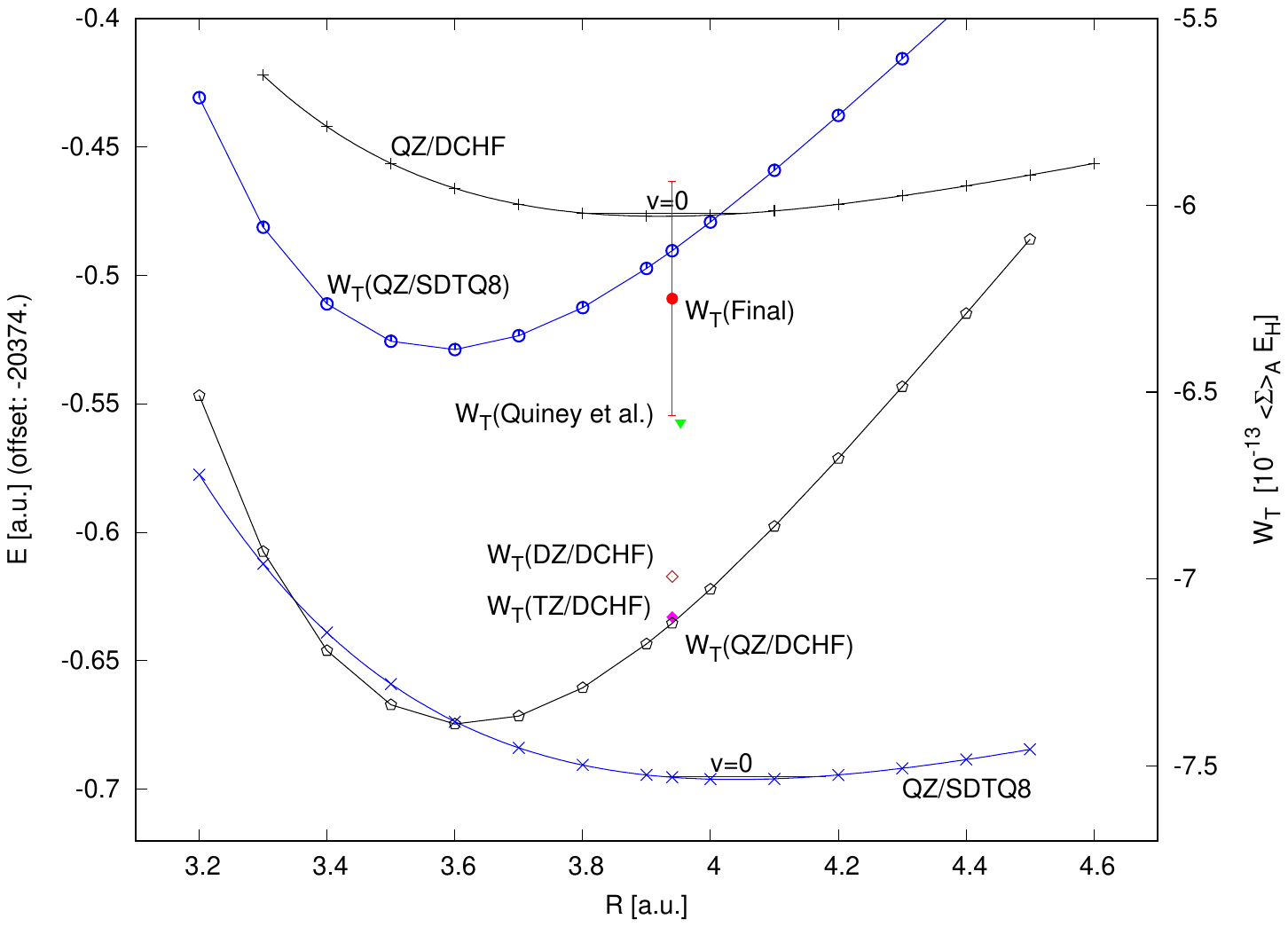}
  \caption{Potential-energy curves (left-hand scale, polynomial-fitted curves) and T-PT interaction constants (right-hand scale,
           connected data points) from selected models and comparison with literature. The final value for $W_T$ is given with
           an uncertainty bar calculated as described in the text.
           \label{FIG:CU_WT}
          }
\end{figure}

Figure \ref{FIG:CU_WT} shows the electronic ground-state potential of TlF along with essential results for the T-PT interaction
at Hartree-Fock and a selected level of correlated theory that yields a result close to the final value for $W_T$.
The single-point results for $W_T$ demonstrate that the value reported by Quiney \etal\ \cite{Quiney_PTodd_PRA1997} is not far from the
present final result but benefitted from a fortuitous error cancellation: More complete basis sets increase the magnitude of
the T-PT interaction and electron correlation effects decrease it. From the functions $W_T(R)$ it is clear that the maximum
polarization of the TlF molecule is achieved around $R = 3.6$ \au\ which is significantly shorter than the equilibrium bond
length $R_e$. This observation has also been made for other ${\cal{P,T}}$-violating interaction constants in other diatomic
molecules \cite{Fleig_DeMille_2021}. Another interesting observation concerns the variation of $W_T$ with molecular vibration.
Between the classical turning points in the respective vibrational ground state $\Delta W_T$ is on the order of $5$\%, but
since $W_T(R)$ is near linear in that regime an averaging over $W_T$ results in a negligible vibrational correction.

\subsubsection{T-PT vs. S-PS interaction}

For the leading closed-shell determinant (active molecular spinors and time-reversed partners given with label $m_j$, the
exact projection of total one-body angular momentum $j$ onto the internuclear axis, and leading atomic character)
$$D_0 = a^{\dagger}_{\text{Tl}(6s_{\frac{1}{2}})}\,  a^{\dagger}_{\text{Tl}\overline{(6s_{\frac{1}{2}})}}\,
        a^{\dagger}_{\text{F}(2p_{\frac{3}{2}})}\, a^{\dagger}_{\text{F}\overline{(2p_{\frac{3}{2}})}}\,
        a^{\dagger}_{\text{F}(2p_{\frac{1}{2}})}\, a^{\dagger}_{\text{F}\overline{(2p_{\frac{1}{2}})}}\,
        a^{\dagger}_{\text{F}(2p_{\frac{1}{2}})}\, a^{\dagger}_{\text{F}\overline{(2p_{\frac{1}{2}})}}\; \rvac$$
in the subspace eigenvector of the CI expansion for the ground state of TlF in the highly-correlated model SDTQ8\_6.5au
the expansion coefficient is found to be $c_{D_0} = -0.9566$ which means that $D_0$ has a weight of about $91.5$\% in
the total wavefunction. As a consequence of this the S-PS-Ne interaction constant in the same state and with the same model 
is non-zero but strongly suppressed, as expected. Calculated in the same way as in Ref. \cite{PhysRevA.96.040502}
the result for $W_S$ is about six orders of magnitude smaller than the corresponding result for HfF$^+$.

\section{Conclusions}
\label{SEC:CONCL}
The recent evidence for Beyond-Standard-Model physics from high-energy collision data in particular in
$b \rightarrow s{\ell}^+ {\ell}^-$ decays -- where $\ell$ represents a lepton and $s$ a {\it{strange}} quark -- 
provides renewed motivation to
interpret EDM searches in closed-shell atoms and molecules in terms of the ${\cal{CP}}$-violating T-PT-ne interaction.
The CeNTREX collaboration \cite{CENTREX_2021} anticipates for the near future a roughly 2500-fold improvement over the 
previous best measurements of the TlF EDM. Combination with the present results for the molecular T-PT interaction constant 
in the TlF molecule will thus allow for obtaining significantly stronger bounds on fundamental CPV parameters, in
particular the semi-leptonic CPV electron-quark coefficients \cite{Fuyuto_RamseyM_leptoq2018}. The latter are generated
by tree-level exchange of the BSM mediator particles, see Fig. (\ref{FIG:SLQEX}).

The present formalism and implementation will in the near future be used to make predictions for T-PT-ne interaction
constants in other diatomic molecules that are perspective systems for next-generation searches of new ${\cal{CP}}$-violation
with orders-of-magnitude improved projected sensitivity, in particular the francium-silver molecule \cite{marc2023candidate}.
\clearpage
\section{Acknowledgments}
\label{SEC:ACK}
I thank Peter Stangl (CERN) for helpful comments.
\bibliographystyle{unsrt}

\begin{thebibliography}{10}

\bibitem{LeptonUniversality_2023}
R.~Aaij \etal\ (LHCb~collaboration).
\newblock {Test of Lepton Universality in $b \rightarrow s{\ell}^+ {\ell}^-$
  Decays}.
\newblock {\em Phys. Rev. Lett.}, 131:051803, 2023.

\bibitem{lepton_univ_viol2022}
LHCb collaboration.
\newblock {Test of lepton universality in beauty-quark decays}.
\newblock {\em Nature Physics}, 18:277, 2022.

\bibitem{AngularAnalysis_BK_decay_2021}
R.~Aaij \etal\ (LHCb~collaboration).
\newblock {Angular Analysis of the in $B^+ \rightarrow K^{*+}{\mu}^+ {\mu}^-$
  Decay}.
\newblock {\em Phys. Rev. Lett.}, 126:161802, 2021.

\bibitem{Becirevic:2018afm}
Damir Bečirević, Ilja Doršner, Svjetlana Fajfer, Nejc Košnik, Darius~A.
  Faroughy, and Olcyr Sumensari.
\newblock {Scalar leptoquarks from grand unified theories to accommodate the
  $B$-physics anomalies}.
\newblock {\em Phys. Rev.}, D98(5):055003, 2018.

\bibitem{Barbieri_compLeptoquark_2017}
R.~Barbieri, C.~W. Murphy, and F.~Senia.
\newblock {B-decay anomalies in a composite leptoquark model}.
\newblock {\em Eur. Phys. J. C}, 77:8, 2017.

\bibitem{Wise_leptoquarks2013}
J.~M. Arnold, B.~Fornal, and M.~B. Wise.
\newblock {Phenomenology of scalar leptoquarks}.
\newblock {\em Phys. Rev. D}, 88:035009, 2013.

\bibitem{Davies_He_scalarfermions1990}
A.~J. Davies and X.-G. He.
\newblock {Tree-level scalar-fermion interactions consistent with the
  symmetries of the standard model}.
\newblock {\em Phys. Rev. D}, 43:225, 1991.

\bibitem{Buchmueller_leptoquarks1987}
W.~Buchm{\"u}ller, R.~R{\"u}ckl, and D.~Wyler.
\newblock {Leptoquarks in lepton-quark collisions}.
\newblock {\em Phys. Lett. B}, 191:442, 1987.

\bibitem{https://doi.org/10.48550/arxiv.2203.08103}
Ricardo Alarcon, Jim Alexander, Vassilis Anastassopoulos, Takatoshi Aoki, Rick
  Baartman, Stefan Baeßler, Larry Bartoszek, Douglas~H. Beck, Franco Bedeschi,
  Robert Berger, Martin Berz, Hendrick~L. Bethlem, Tanmoy Bhattacharya, Michael
  Blaskiewicz, Thomas Blum, Themis Bowcock, Anastasia Borschevsky, Kevin Brown,
  Dmitry Budker, Sergey Burdin, Brendan~C. Casey, Gianluigi Casse, Giovanni
  Cantatore, Lan Cheng, Timothy Chupp, Vince Cianciolo, Vincenzo Cirigliano,
  Steven~M. Clayton, Chris Crawford, B.~P. Das, Hooman Davoudiasl, Jordy
  de~Vries, David DeMille, Dmitri Denisov, Milind~V. Diwan, John~M. Doyle,
  Jonathan Engel, George Fanourakis, Renee Fatemi, Bradley~W. Filippone,
  Victor~V. Flambaum, Timo Fleig, Nadia Fomin, Wolfram Fischer, Gerald
  Gabrielse, R.~F.~Garcia Ruiz, Antonios Gardikiotis, Claudio Gatti, Andrew
  Geraci, James Gooding, Bob Golub, Peter Graham, Frederick Gray, W.~Clark
  Griffith, Selcuk Haciomeroglu, Gerald Gwinner, Steven Hoekstra, Georg~H.
  Hoffstaetter, Haixin Huang, Nicholas~R. Hutzler, Marco Incagli, Takeyasu~M.
  Ito, Taku Izubuchi, Andrew~M. Jayich, Hoyong Jeong, David Kaplan, Marin
  Karuza, David Kawall, On~Kim, Ivan Koop, Wolfgang Korsch, Ekaterina
  Korobkina, Valeri Lebedev, Jonathan Lee, Soohyung Lee, Ralf Lehnert, Kent
  K.~H. Leung, Chen-Yu Liu, Joshua Long, Alberto Lusiani, William~J. Marciano,
  Marios Maroudas, Andrei Matlashov, Nobuyuki Matsumoto, Richard Mawhorter,
  Francois Meot, Emanuele Mereghetti, James~P. Miller, William~M. Morse, James
  Mott, Zhanibek Omarov, Luis~A. Orozco, Christopher~M. O'Shaughnessy, Cenap
  Ozben, SeongTae Park, Robert~W. Pattie, Alexander~N. Petrov, Giovanni~Maria
  Piacentino, Bradley~R. Plaster, Boris Podobedov, Matthew Poelker, Dinko
  Pocanic, V.~S. Prasannaa, Joe Price, Michael~J. Ramsey-Musolf, Deepak
  Raparia, Surjeet Rajendran, Matthew Reece, Austin Reid, Sergio Rescia, Adam
  Ritz, B.~Lee Roberts, Marianna~S. Safronova, Yasuhiro Sakemi, Philipp
  Schmidt-Wellenburg, Andrea Shindler, Yannis~K. Semertzidis, Alexander
  Silenko, Jaideep~T. Singh, Leonid~V. Skripnikov, Amarjit Soni, Edward
  Stephenson, Riad Suleiman, Ayaki Sunaga, Michael Syphers, Sergey Syritsyn,
  M.~R. Tarbutt, Pia Thoerngren, Rob G.~E. Timmermans, Volodya Tishchenko,
  Anatoly~V. Titov, Nikolaos Tsoupas, Spyros Tzamarias, Alessandro Variola,
  Graziano Venanzoni, Eva Vilella, Joost Vossebeld, Peter Winter, Eunil Won,
  Anatoli Zelenski, Tanya Zelevinsky, Yan Zhou, and Konstantin Zioutas.
\newblock Electric dipole moments and the search for new physics, 2022.

\bibitem{Barr_eN-EDM_Atoms_1992}
S.~M. Barr.
\newblock ${T}$- and ${P}$-odd electron-nucleon interactions and the electric
  dipole moments of large atoms.
\newblock {\em Phys. Rev. D}, 45:4148, 1992.

\bibitem{He_McKellar_eNinteractions_1992}
X.-G. He, B.~H.~J. McKellar, and S.~Pakvasa.
\newblock {$CP$-violating electron-nucleus interactions in multi-Higgs doublet
  and leptoquark models}.
\newblock {\em Phys. Lett. B}, 283:348, 1992.

\bibitem{Cairncross_Ye_NatPhys2019}
W.~B. Cairncross and J.~Ye.
\newblock {Atoms and molecules in the search for time-reversal symmetry
  violation}.
\newblock {\em Nature Physics}, 1:510, 2019.

\bibitem{ginges_flambaum2004}
J.~S.~M. Ginges and V.~V. Flambaum.
\newblock Violations of fundamental symmetries in atoms and tests of
  unification theories of elementary particles.
\newblock {\em Phys. Rep.}, 397:63, 2004.

\bibitem{Heckel_Hg_PRL2016}
B.~Graner, Y.~Chen, E.~G. Lindahl, and B.~R. Heckel.
\newblock Reduced {L}imit on the {P}ermanent {E}lectric {D}ipole {M}oment of
  {$^{199}${H}g}.
\newblock {\em Phys. Rev. Lett.}, 116:161601, 2016.

\bibitem{CENTREX_2021}
O.~Grasdijk, O.~Timgren, J.~Kastelic, T.~Wright, S.~Lamoreaux, D.~DeMille,
  K.~Wenz, M.~Aitken, T.~Zelevinsky, T.~Winick, and D.~Kawall.
\newblock {CeNTREX: A new search for time-reversal symmetry violation in the
  {$^{205}$Tl} nucleus}.
\newblock {\em Quantum Sci. Technol.}, 6:044007, 2021.

\bibitem{Hubert_Fleig_2022}
M.~Hubert and T.~Fleig.
\newblock {Electric dipole moments generated by nuclear Schiff moment
  interactions: A reassessment of the atoms {$^{129}$Xe} and {$^{199}$Hg} and
  the molecule {$^{205}$TlF} }.
\newblock {\em Phys. Rev. A}, 106:022817, 2022.

\bibitem{Flambaum-Dzuba_TranPRA2020}
V.~V. Flambaum, V.~A. Dzuba, and H.~B.~Tran Tan.
\newblock {Time- and parity-violating effects of the nuclear Schiff moment in
  molecules and solids}.
\newblock {\em Phys. Rev. A}, 101:042501, 2020.

\bibitem{Proctor_nuclear1950}
J.~A. Bounds, C.~R. Bingham, H.~K. Carter, G.~A. Leander, R.~L. Mlekodaj, E.~H.
  Spejewski, and Jr. W.~M.~Fairbank.
\newblock {On the Magnetic Moments of Tl$^{203}$, Tl$^{205}$, Sn$^{115}$,
  Sn$^{117}$, Sn$^{119}$, Cd$^{111}$, Cd$^{113}$, and Pb$^{207}$}.
\newblock {\em Phys. Rev.}, 79:35, 1950.

\bibitem{Bounds_Tl_nuclear1987}
J.~A. Bounds, C.~R. Bingham, H.~K. Carter, G.~A. Leander, R.~L. Mlekodaj, E.~H.
  Spejewski, and Jr. W.~M.~Fairbank.
\newblock {Nuclear structure of light thallium isotopes as deduced from laser
  spectroscopy on a fast atom beam}.
\newblock {\em Phys. Rev. C}, 36:2560, 1987.

\bibitem{Budincevic_diss}
Ivan Budinčević.
\newblock {\em Nuclear structure studies of rare francium isotopes using
  Collinear Resonance Ionization Spectroscopy (CRIS)}.
\newblock Dissertation, Arenberg Doctoral School, Faculty of Science, KU Leuven
  (Belgium), 2015.

\bibitem{Flambaum-Dzuba_PRA2020}
V.~V. Flambaum and V.~A. Dzuba.
\newblock {E}lectric dipole moments of atoms and molecules produced by enhanced
  nuclear schiff moments.
\newblock {\em Phys. Rev. A}, 101:042504, 2020.

\bibitem{Coveney_JPB1983}
P.~V. Coveney and P.~G.~H. Sandars.
\newblock Parity- and time-violating interactions in thallium fluoride.
\newblock {\em J. Phys. B: At. Mol. Opt. Phys.}, 16:3727, 1983.

\bibitem{Quiney_PTodd_PRA1997}
H.~M. Quiney, J.~K. L{\ae}rdahl, T.~Saue, and K.~F{\ae}gri Jr.
\newblock {$Ab~initio$ Dirac-Hartree-Fock calculations of chemical properties
  and $PT$-odd effects in thallium fluoride}.
\newblock {\em Phys. Rev. A}, 57:920, 1998.

\bibitem{Chupp_Ramsey_Global2015}
T.~Chupp and M.~Ramsey-Musolf.
\newblock {E}lectric dipole moments: {A} global analysis.
\newblock {\em Phys. Rev. C}, 91:035502, 2015.

\bibitem{FleigJung_JHEP2018}
T.~Fleig and M.~Jung.
\newblock {M}odel-independent determinations of the electron {EDM} and the role
  of diamagnetic atoms.
\newblock {\em J. High Energy Phys.}, 07:012, 2018.

\bibitem{Fleig_Jung_Xe_2021}
T.~Fleig and M.~Jung.
\newblock {$P,T$-Odd Interactions in Atomic {$^{129}$Xe} and Phenomenological
  Applications}.
\newblock {\em Phys. Rev. A}, 103:012807, 2021.

\bibitem{fleig_gasci}
T.~Fleig, J.~Olsen, and C.~M. Marian.
\newblock The generalized active space concept for the relativistic treatment
  of electron correlation. {I}. {K}ramers-restricted two-component
  configuration interaction.
\newblock {\em J. Chem. Phys.}, 114:4775, 2001.

\bibitem{fleig_gasci2}
T.~Fleig, J.~Olsen, and L.~Visscher.
\newblock The generalized active space concept for the relativistic treatment
  of electron correlation. {II}: {L}arge-scale configuration interaction
  implementation based on relativistic 2- and 4-spinors and its application.
\newblock {\em J. Chem. Phys.}, 119:2963, 2003.

\bibitem{Fleig_PRA2019}
T.~Fleig.
\newblock {P,T}-odd tensor-pseudotensor interactions in atomic {$^{199}$Hg} and
  {$^{225}$Ra}.
\newblock {\em Phys. Rev. A}, 99:012515, 2019.

\bibitem{Fleig_Skripnikov2020}
T.~Fleig and L.~V. Skripnikov.
\newblock {P,T-Violating and Magnetic Hyperfine Interactions in Atomic
  Thallium}.
\newblock {\em Symmetry}, 12:498, 2020.

\bibitem{dyall_p}
K.~G. Dyall.
\newblock Relativistic and nonrelativistic finite nucleus optimized triple-zeta
  basis sets for the 4p, 5p and 6p elements.
\newblock {\em Theoret. Chem. Acc.}, 108:335, 2002.

\bibitem{dyallBi2}
K.~G. Dyall.
\newblock {Relativistic and nonrelativistic finite nucleus optimized triple
  zeta basis sets for the 4p, 5p and 6p elements}.
\newblock {\em Theoret. Chem. Acc.}, 109:284, 2003.

\bibitem{4p-basis-dyall-2}
K.~G. Dyall.
\newblock Relativistic double-zeta, triple-zeta, and quadruple-zeta basis sets
  for the 4p, 5p and 6p elements.
\newblock {\em Theoret. Chim. Acta}, 115:441, 2006.

\bibitem{EMSL-basis2019}
B.~P. Pritchard, D.~Altarawy, B.~Didier, T.~D. Gibson, and T.~L. Windus.
\newblock {A New Basis Set Exchange: An Open, Up-to-date Resource for the
  Molecular Sciences Community.}
\newblock {\em J. Chem. Inf. Model.}, 59:4814, 2019.

\bibitem{DIRAC_JCP}
T.~Saue, R.~Bast, A.~S.~P. Gomes, H.~J.~Aa. Jensen, L.~Visscher, I.~A. Aucar,
  R.~Di Remigio, K.~G. Dyall, E.~Eliav, E.~Fasshauer, T.~Fleig, L.~Halbert,
  E.~Donovan Hedeg{\aa}rd, B.~Helmich-Paris, M.~Ilia{\v{s}}, C.~R. Jacob,
  S.~Knecht, J.~K. Laerdahl, M.~L. Vidal, M.~K. Nayak, M.~Olejniczak,
  J.~M.~Haugaard Olsen, M.~Pernpointner, B.~Senjean, A.~Shee, A.~Sunaga, and
  J.~N.~P. van Stralen.
\newblock {The DIRAC code for relativistic molecular calculations}.
\newblock {\em J. Phys. Chem.}, 152:204104, 2020.

\bibitem{knecht_luciparII}
S.~Knecht, H.~J.~Aa. Jensen, and T.~Fleig.
\newblock {L}arge-{S}cale {P}arallel {C}onfiguration {I}nteraction. {II}.
  {T}wo- and four-component double-group general active space implementation
  with application to {B}i{H}.
\newblock {\em J. Chem. Phys.}, 132:014108, 2010.

\bibitem{Barrett_Mandel_1957}
A.~H. Barrett and M.~Mandel.
\newblock {Microwave Spectra of the Tl, In, and Ga Monohalides}.
\newblock {\em Phys. Rev.}, 109:1572, 1957.

\bibitem{ChoSangsterHinds_TlF_PRA1991}
D.~Cho, K.~Sangster, and E.~A. Hinds.
\newblock {Search for time-reversal-symmetry violation in thallium fluoride
  using a jet source}.
\newblock {\em Phys. Rev. A}, 44:2783, 1991.

\bibitem{PhysRevA.95.022504}
Timo Fleig.
\newblock {T}a{O}$^{+}$ as a candidate molecular ion for searches of physics
  beyond the standard model.
\newblock {\em Phys. Rev. A}, 95:022504, Feb 2017.

\bibitem{singh_CPodd-Ra_PRA2015}
Y.~Singh and B.~K. Sahoo.
\newblock {Electric dipole moment of $^{225}$Ra due to P- and T -violating weak
  interactions}.
\newblock {\em Phys. Rev. A}, 92:022502, 2015.

\bibitem{Fleig_DeMille_2021}
T.~Fleig and D.~DeMille.
\newblock {Theoretical aspects of radium-containing molecules amenable to
  assembly from laser-cooled atoms for new physics searches}.
\newblock {\em New J. Phys.}, 23:113039, 2021.
\newblock Erratum: {\em{New J. Phys.}}, 23:113039, 2021.

\bibitem{PhysRevA.96.040502}
Timo Fleig.
\newblock $\mathcal{P},\mathcal{T}$-odd and magnetic hyperfine-interaction
  constants and excited-state lifetime for {H}f{F}$^+$.
\newblock {\em Phys. Rev. A}, 96:040502, Oct 2017.

\bibitem{Fuyuto_RamseyM_leptoq2018}
K.~Fuyuto, M.~J. Ramsey-Musolf, and T.~Shen.
\newblock {Electric Dipole Moments from CP-Violating Scalar Leptoquark Models}.
\newblock {\em Phys. Lett. B}, 788:52, 2019.

\bibitem{marc2023candidate}
Aurélien Marc, Mickaël Hubert, and Timo Fleig.
\newblock Candidate molecules for next-generation searches of hadronic
  charge-parity violation, 2023.
\newblock arXiv physics.atom-ph (2023) 2309.11633.

\end{thebibliography}

\newcommand{\Aa}[0]{Aa}


\end{document}